# Application of a two-temperature model for the investigation of the periodic structure formation on Si surface in femtosecond laser interactions

T. JY. DERRIEN[a*], T. SARNET[a], M. SENTIS[a], T. E. ITINA[a,b]

[a]Laboratoire Lasers, Plasmas et Procédés Photoniques, UMR CNRS 6182/Université de la Méditerranée, 163 avenue de Luminy, 13288 Marseille, France
[b]Laboratoire Hubert Curien, UMR CNRS 5516/Université de Lyon, 18 rue Benoît Lauras, Bat. F, 42000, Saint-Etienne, France

We consider the case of surface irradiation by a small number of femtosecond laser shots leading to the formation of surface ripples. To explain this effect, we propose a numerical model that accounts for the following processes: (i) interference of the laser irradiation with an electromagnetic surface wave propagating on a silicon sample; (ii) free carrier formation and laser energy absorption; (iii) energy relaxation and electron-phonon coupling. We perform numerical calculations taking into account the interference of a surface wave with laser; and present the obtained simulation results in order to explain formation mechanisms of the experimentally observed patterns.

## 1. Introduction

Ultrashort pulse laser interactions have many applications in different areas, such as laser micro-machining, material analysis, nano-particle formation, etc. In particular, femtosecond lasers can be used to form surface periodic structures (LIPSS). These structures can find an application in the development of photovoltaic cells (due to enhanced absorption, ex: black silicon [1,2]). In particular, previous experiments demonstrated that, with small laser fluence and small number of laser shots, near wavelength spaced periodic structures called "ripples" appear on the surface of laser-irradiated material. When larger number of shots is applied, elongated structures were shown to grow, providing good absorption properties of light in ultraviolet and visible wavelengths [3]. This structure formation was observed for different target materials (metals [4], semiconductors and dielectrics [5,6]). Designing the best shape for increasing absorption on semiconductors (photovoltaic cells, photosensors, …) requires a better understanding of the growth of spikes and ripples in femtosecond laser interactions [7].

Previously, numerous investigations, both experimental and theoretical, considered LIPSS [1,8-12]. For instance, the model of Sipe considers electrodynamic interactions of laser wave with defects of materials, modifying electromagnetic fields, polarization, and energy absorption of the target [13]. The LIPSS formation process was furthermore described by Kuramoto-Sivashinsky equation [14]. A key idea of this model is in the formation of self-organized structures [7]. In addition, to explain femtosecond laser interaction leading to the appearance of the so-called "High Spatial Frequency LIPSS", or HSFL, a model based on the second harmonic generation was proposed [15]. Furthermore, other explanations involved the surface waves on metals surfaces (surface plasmons [16]), or surface instabilities [17]. However, fused dielectric materials have a transient metallic behaviour [18] after free carrier generation by one or multiphoton ionization.

In this paper, we propose a two-dimensional model that accounts for the interference of laser wave with another electro-magnetic wave ("surface" wave). The model provides some insights on the density profile of the laser generated free-carriers and temperatures.

## 2. Model and calculation details

To study the mechanisms of femtosecond laser interaction with silicon, we use a *two temperature model* (TTM) describing electron and lattice thermal behavior [19]. In our equation system, the surface is at ($x$; $z$=0) and z is the depth. Solid to liquid phase transition is considered by using the enthalpy, $H$, while lattice temperature equals fusion temperature of silicon. We solve the following set of equations for the electrons (1) and the lattice for bulk Silicon

$$\begin{cases} C_e \dfrac{\partial T_e(t,x,z)}{\partial t} = \nabla(\kappa_e(T_e).\nabla T_e) - \gamma_{ei}(T_e - T_i) + S(t,x,z) \\ C_i \dfrac{\partial T_i(t,x,z)}{\partial t} = \nabla(\kappa_i(T_i)\nabla T_i) + \gamma_{ei}(T_e - T_i) \end{cases}$$

$$(1)$$

where $T_e$ and $T_i$ are electron and lattice temperature; $C_e = \dfrac{3}{2} k_B n_e$ is the specific heat capacity of electrons; $\kappa_e = \dfrac{2 k_B \mu_e n_e T_e}{e}$ is the heat conductivity for electrons [20]; $k_B$ is the Boltzmann constant; $\mu_e$ is the electron mobility; $n_e$ is the free carrier density; $e$ is the charge of electron; $C_i$ and $\kappa_i$ are taken from [21]. $\gamma_{ei} = \dfrac{C_e}{\tau_E}$ describes the electron-lattice coupling; $\tau_E = 10^{-13} s$ is the electron-phonon collision time [21].

Equation (1) is solved in two dimensions (cartesian in depth and surface) by using an implicit scheme and *Forward Elimination – Backward Substitution* algorithm. Initial condition is $T_e = T_i = 300 K$, and boundary conditions are Neumann's conditions.

We consider the ionization processes in silicon by using the free carrier balance equation (2) that accounts one and two photon ionization, Auger recombination, and impact ionization [20]

$$\frac{\partial n_e}{\partial t} = \frac{A \sigma_1 I(t,x,z)}{\hbar \omega_{laser}} + \frac{A^2 \sigma_2 I^2(t,x,z)}{2 \hbar \omega_{laser}} + \delta(T_e) n_e - \gamma_{Auger} n_e^3 \tag{2}$$

where $n_e [m^{-3}]$ is the free carrier density (electrons-holes), $A = \dfrac{1-R}{\cos \theta_i}$ is the absorptivity of laser wave on silicon surface, $\sigma_1 [m^{-1}]$ is the one-photon-ionization rate, $\sigma_2 [m.W^{-1}]$ is the two-photon-ionization rate, $\delta(T_e) [s^{-1}]$ is the impact ionization (electronic avalanche) rate, and $\gamma_{Auger} [m^6 . s^{-1}]$ is the Auger recombination rate of free carriers.

Numerical solution is obtained by using an explicit Runge-Kutta 4th order scheme.

The reflectivity, $R$, and the energy absorption coefficient, $\alpha_{abs}$, are deduced from dielectric function (3) based on the Drude's model [20].

$$\left\{ \begin{array}{l} \varepsilon = \varepsilon_1 + i.\varepsilon_2, \\[2mm] \varepsilon_1 = 1 + \left( \mathrm{Re}(\varepsilon_g) - 1 \right) \left( 1 - \dfrac{n_e}{n_0} \right) - \dfrac{n_e}{n_{critical}} \left[ 1 + \left( \dfrac{\omega_{laser}}{\Gamma} \right)^{-2} \right]^{-1}, \\[4mm] \varepsilon_2 = \dfrac{n_e}{n_{critic}} \left[ \left( \dfrac{\omega_{laser}}{\Gamma} \right) \left( 1 + \left( \dfrac{\omega_{laser}}{\Gamma} \right)^{-2} \right) \right]^{-1} + \mathrm{Im} \left( \varepsilon_g \left( 1 - \dfrac{n_e}{n_0} \right) \right), \end{array} \right. \tag{3}$$

where $n_{cr}$ is the critical density needed for optical breakdown in silicon, $n_{cr} = \dfrac{m_e \varepsilon_0 \omega_{laser}^2}{e^2}$; $n_0$ is the valence band density, $\varepsilon_g$ is the complex dielectric constant of unexcited material [20].

The initial condition for the free carrier (electrons or holes) density is

$$n_e(t = t_{\min}) = \left( \frac{2^{\frac{5}{2}} \left( m_e \pi k_B \right)^{\frac{3}{2}}}{h^3} \right) T_e^{\frac{3}{2}} (t = t_{\min}) e^{-\frac{E_g}{2 k_B T_e}},$$

here $E_g$ is the silicon band gap; $h$ is the Planck constant [22].

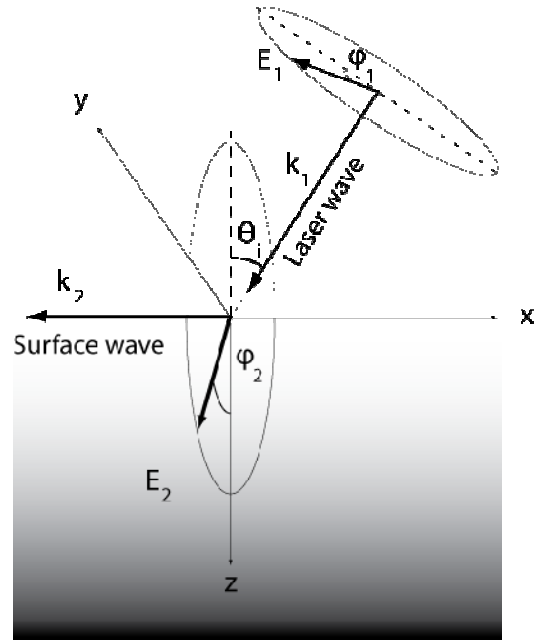

*Fig. 1. Schematics of the two-wave interaction in the cartesian frame.*

To calculate the energy source considering the interference effect, we use the plane wave expressions with the following electric fields

$$\overrightarrow{E_1} = \overrightarrow{E_{1,0}}(\theta_i, \phi_1) \cos(\omega_1 t + \psi_1),$$
$$\overrightarrow{E_2} = \overrightarrow{E_{2,0}}(\phi_2) \cos(\omega_2 t + \psi_2),$$

where $\overrightarrow{E_1}$ is the laser wave; $\overrightarrow{E_2}$ is the surface wave (the origin of the surface wave is explained in the next section); $\phi_1$ and $\phi_2$ govern electric field polarization;

$\omega_1 = 2\pi \dfrac{c}{\lambda_1}$ and $\omega_2 = 2\pi \dfrac{c}{\lambda_2}$ are laser and surface wave pulsations where $\lambda_1$ is laser wavelength and $\lambda_2$ is surface wave wavelength; $\psi_1 = -\vec{k_1}.\vec{r}$ and $\psi_2 = -\vec{k_2}.\vec{r}$ are wave phases where

$$\vec{k_1} = \frac{\omega_1}{c}\begin{pmatrix} -\sin\theta_i \\ 0 \\ \cos\theta_i \end{pmatrix}, \ \vec{k_2} = \frac{\omega_2}{c}\begin{pmatrix} -1 \\ 0 \\ 0 \end{pmatrix}, \ \vec{r} = \begin{pmatrix} x \\ y \\ z \end{pmatrix}.$$

We also define

$$\vec{E_{1,0}}(\theta_i, \phi_1) = \begin{pmatrix} -\cos\theta_i \cos\phi_1 \\ \sin\phi_1 \\ -\sin\theta_i \cos\phi_1 \end{pmatrix},$$

$$\vec{E_{2,0}}(\theta_i, \phi_2) = \begin{pmatrix} 0 \\ \sin\phi_2 \\ \cos\phi_2 \end{pmatrix}$$

as shown on
Fig. *1*. The total intensity (Poynting vector) on target is

$$\vec{I}(t,x,z) = \frac{1}{2}c\varepsilon_0 \frac{\vec{k_1}+\vec{k_2}}{\|\vec{k_1}+\vec{k_2}\|}\Big[E_{1,0}{}^2\cos^2(\omega_1 t + \psi_1) + E_{2,0}{}^2\cos^2(\omega_2 t + \psi_2) + \ldots$$
$$\ldots + 2E_{1,0}E_{2,0}\cos(\omega_1 t + \psi_1)\cos(\omega_2 t + \psi_2)\cos(\phi_1 - \phi_2)(1 - \sin\theta_i)\Big]$$

We assume that the beam is Gaussian (spot size $\Delta_{spot}$ Full Width at Half Maximum – FWHM). We also chose a Gaussian distribution for the temporal profile ($\tau$ at FWHM).

$$I(t,x,z) = \left\|\langle\vec{I}(t,x;z=0)\rangle_{T_1}\right\| e^{-\frac{1}{2}\left(\frac{x}{\sigma_x}\right)^2} e^{-\frac{1}{2}\left(\frac{t}{\sigma_\tau}\right)^2} e^{-\alpha_{abs}z}$$

$, \ \sigma_x = \dfrac{\Delta_{spot}}{2\sqrt{2\ln(2)}}, \ \sigma_\tau = \dfrac{\tau}{2\sqrt{2\ln(2)}}$

To calculate the laser intensity, we take the average value over the laser period $T_1 = \dfrac{\lambda_1}{c}$ ., If $\lambda_1 = \lambda_2$,

$$\left\|\langle\vec{I}(t,x;z=0)\rangle_{T_1}\right\| = \frac{1}{2}c\varepsilon_0\left[\frac{E_{1,0}{}^2}{2} + \frac{E_{2,0}{}^2}{2} + E_{1,0}E_{2,0}\cos(\psi_1 - \psi_2)\right]$$

Although, if $\lambda_1 \neq \lambda_2$, average on laser period is

$$\left\|\langle\vec{I}(t,x;z=0)\rangle_{T_1}\right\| = \frac{1}{2}c\varepsilon_0 \left[ \begin{array}{l} \dfrac{E_{1,0}{}^2}{2} + \dfrac{E_{2,0}{}^2}{8\pi}\left(\dfrac{\omega_1}{\omega_2}\right)\left(\sin\left(4\pi\dfrac{\omega_2}{\omega_1} + 2\psi_2\right) + 4\pi\dfrac{\omega_2}{\omega_1} - \sin(2\psi_2)\right) \\[3mm] + \dfrac{E_{1,0}E_{2,0}}{2\pi}\omega_1 \left[ \begin{array}{l} \dfrac{\sin\left(\dfrac{2\pi}{\omega_1}(\omega_1 - \omega_2) + (\psi_1 - \psi_2)\right)}{\omega_1 - \omega_2} + \dfrac{\sin\left(\dfrac{2\pi}{\omega_1}(\omega_1 + \omega_2) + (\psi_1 + \psi_2)\right)}{\omega_1 + \omega_2} \\[3mm] - \dfrac{\sin(\psi_1 + \psi_2)}{\omega_1 + \omega_2} - \dfrac{\sin(\psi_1 - \psi_2)}{\omega_1 - \omega_2} \end{array} \right] \end{array} \right]$$

To observe the persistency of a simple interference pattern in free carrier density and temperatures, we consider the interference between the incident and the reflected wave on surface by taking $\lambda_1 = \lambda_2 = 800\,nm$.

The laser fluence $f$ absorbed by the silicon target is given by the direct interband absorption, one photon ionization, Auger recombination, and impact ionization [20].

Therefore, the source term for the TTM 2D equation (1) is:

$$S(t,x,z) = A\alpha_{abs}I(t,x,z) + (\hbar\omega_{laser} - E_{gap})\frac{A\sigma_1 I(t,x,z)}{\hbar\omega_{laser}} + \ldots$$
$$+ E_{gap}\gamma_{Auger}n_e{}^3 - E_{gap}\delta(T_e)n_e - \frac{3}{2}k_B T_e \frac{\partial n_e}{\partial t}$$
(4)

The calculation parameters used in the TTM model and the free carrier balance equation (2) are described in Table 1.

*Table1. Parameters used to solve the TTM equation (1), free carrier equation (2) and compute the dielectric function (3).*

| Parameter | Value | Unit | Reference |
|-----------|-------|------|-----------|
| $C_e$ | $\dfrac{3}{2}k_B n_e$ | $J.m^{-3}.K^{-1}$ | [20] |
| $\kappa_e$ | $\dfrac{2k_B{}^2 \mu_e n_e T_e}{e}, \mu_e = 0.015\, m^2.V^{-1}.s^{-1}$ | $W.m^{-2}.K^{-1}$ | [20] |
| $C_i$ | $10^6\left(1.978 + 3.54\times10^{-4}T_i - 3.68T_i^{-2}\right)$ | $J.m^{-3}.K^{-1}$ | [21] |
| $\kappa_i$ | $1585 T_i^{-1.23}$ | $W.m^{-2}.K^{-1}$ | [21] |
| $\gamma_{ei}$ | $\dfrac{C_e}{\tau_E}, \tau_E = 10^{-13}\, s$ | $W.m^{-3}.K^{-1}$ | [21] |
| $\sigma_1$ | $1.021\times10^5$ | $m^{-1}$ | [20] |
| $\sigma_2$ | $10^{-10}$ | $m.W^{-1}$ | [20] |
| $\delta(T_e)$ | $3.6\times10^{10}\, e^{\frac{1,5E_{gap}}{k_B T_e}}$ | $s^{-1}$ | [21] |
| $\gamma_{Auger}$ | $3.8\times10^{-43}$ | $m^6.s^{-1}$ | [21] |
| $E_{gap}$ | $1.2$ | $eV$ | [20] |
| $\varepsilon_g$ | $13.4 + i0.048$ | - | [20] |
| $n_0$ | $5\times10^{28}$ | $m^{-3}$ | [20] |
| $\Gamma$ | $2.5\times10^{15}$ | $m^{-3}.s^{-1}$ | [20] |

## 3. Results and discussion

The developed model describing both material ionization and the electron-phonon coupling is used to study femtosecond interactions with a silicon target (mono-crystalline sample). A series of calculations is performed to examine the density of free carriers and electron temperature on surface under interfered laser irradiation.

Fig. 2 shows the effect of the laser pulse on electronic temperature and free carrier density distributions on surface, at 1 ps after the maximum of the femtosecond laser pulse. Laser fluence is 1 J.cm$^{-2}$. We observe that the temperature distribution follows the laser intensity profile. The radial distribution of free carrier density, $n_e$, shows that a saturation regime has been reached. The obtained shape of the $n_e(x)$ distribution can be explained by the saturation in laser light absorption that occurs as soon as the density of free carriers reaches the critical plasma density [23]

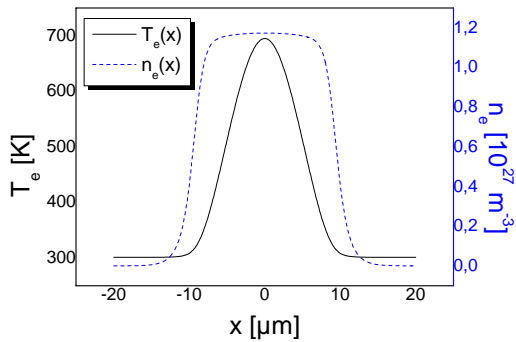

*Fig. 2. Calculated distribution of electronic temperature, $T_e$, and of free carrier density, $n_e$, at the laser-irradiated surface without surface wave. The presented profiles are obtained at a time delay t=1 ps after the maximum of the Gaussian laser pulse, for pulse width of 100 fs, at 800 nm, for single laser pulse. Spot diameter (FWMH) is 10 μm; and laser fluence is 1 J.cm$^{-2}$. Sample is 40 μm large, and 10 μm depth.*

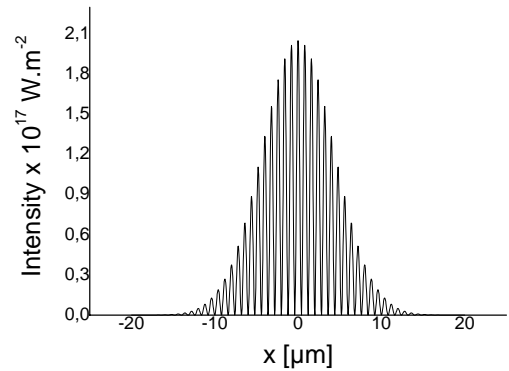

*Fig. 3. Radial intensity distribution obtained at the laser irradiated surface with the interference between the laser pulse and the surface wave. Calculation results are obtained at t=0 ps, z=0 μm, with $\lambda_1 = \lambda_2 = 800$ nm.*

In Fig. 3, we plotted the radial intensity distribution produced at $t=0$ ps on the laser-irradiated surface for two waves. The effect of interference described in section 2 is taken into account. Here, the maximum intensity is $2.10^{17}$ W.m$^{-2}$ corresponds to 1 J.cm$^{-2}$ of total fluence. The shape of the laser profile is Gaussian and the spot diameter (FWMH) is 10 μm. The spacing between oscillations is $\Lambda = 800$ nm. Here, we set $\lambda_1 = \lambda_2 = 800$ nm to simplify the calculations of the average of the Poynting's vector over the oscillation period. We choose electric field amplitudes $E_{1,0}$ and $E_{2,0}$ to provide a good contrast for observing oscillations. Maximum interference amplitude is normalized to the fluence of 1 J.cm$^{-2}$.

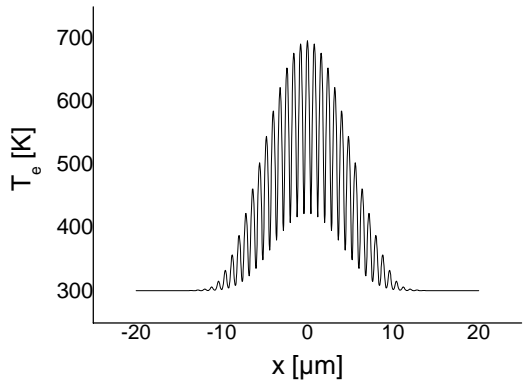

*Fig. 4. Calculated radial distribution of the electronic temperature at 1 ps time delay from maximum laser pulse oscillation. (z=0 μm).*

Fig. 4 shows the radial distribution of the electronic temperature at the surface at $t=1$ ps after laser maximum intensity. At this time, electron energy is transferred to the lattice, and diffuses along both the $x$ and the $z$ axis. The observed shape of the electron temperature profile is affected by both the interference and the thermal diffusion processes.

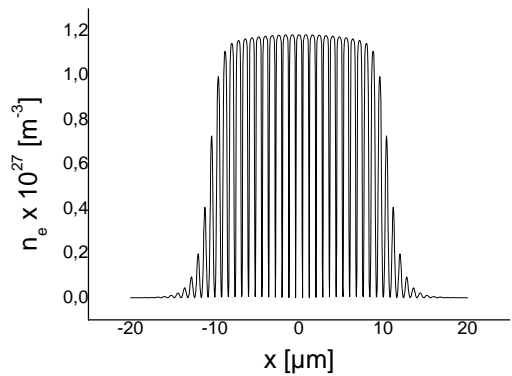

*Fig. 5. Radial distribution of free carrier density $n_e$ at z=0, t=1 ps after the maximum intensity of laser pulse. Here the interference with the surface wave is considered. The laser spot diameter (FWMH) is 10 μm, total fluence is 1 J.cm$^{-2}$, wavelengths of laser and surface waves are 800 nm.*

Fig. 5 shows the radial profile of free carrier density obtained at the surface in the calculations with the effect of the interference described above. The result demonstrates that the interference pattern is conserved in the free carrier density surface profile at a delay of 1 ps after maximum irradiance of the laser pulse. In our simulations, the interference pattern is not erased by thermal diffusion. An estimation of thermal diffusion time for free carrier to transport on 800 nm confirms that this pattern can exist for several picoseconds. Diffusion length at $t=1$ ps after the maximum of intensity is $L_{diff} \approx 190$ nm which is not sufficient to make interferences disappear at this time. At $t=5$ ps after maximum intensity, thermal diffusion length is estimated to 375 nm. These orders of magnitude show that the interference pattern of the free carrier density can last several picoseconds, and it is possible to observe the same pattern in the lattice temperature distribution by electron-phonon coupling, described by the $\gamma_{ei}$ coupling term in the equation (1).

Our results demonstrate that the interference of laser wave with a surface wave produces oscillations in intensity on the surface of the silicon sample. These oscillations are also observed in the electron temperature distribution, and in the free carrier density distribution on surface with a 800 nm periodicity. We also observed a saturation regime in free carrier density, which demonstrates that free carrier generation is limited. The threshold fluence for observing saturation is 0.12 J.cm$^{-2}$.

By modulating the spatial shape of intensity, we demonstrate that it is possible to make interferences in the free carrier density, which can be observed for several picoseconds. We precise that several processes can be involved in the formation of the second wave:

-      The interactions of laser wave with surface defects can lead to the diffraction and the interference that result in the surface wave formation. This explanation was proposed in Sipe et al. paper [13];

-      The observed free carrier density shows gradients which can lead to electronic plasma waves until flattening of theses gradients or recombination of free carriers with the lattice. This process may lead to the formation of the surface plasmon wave.

## 4. Conclusions

In this paper, we have developed a model to investigate the laser interaction with a silicon surface taking into account the interferences between the laser electric field and a surface wave. We have obtained the radial distribution of free carrier density, and the radial electron temperature distribution.

Carrier density profiles show the following:

-      If laser wave interferes with a surface wave, the resulted intensity oscillations appear in the free carrier density profile;

-      The period of the oscillations corresponds to the diffraction period;

-      Thermal diffusion does not erase the free carrier density oscillations after a 1 ps delay ;

-      A saturation regime is observed in the free carrier density for a fluence greater than 0.12 J.cm$^{-2}$.

-      Electron temperature oscillations can be transmitted to the lattice temperature by the process of electron-phonon diffusion.

More calculations will provide some results on the possibility to print the subsystem thermal pattern on the lattice temperature distribution with the electron-phonon diffusion process.

These results indicate that the interferences with the surface wave can be considered in the explanation of femtosecond LIPSS growth process. Such an effect can explain why oscillations are kept in memory of the material under nanotexturing femtosecond laser irradiation

### Acknowledgments

This work is funded by French Ministry of Research under Priority Topic Ph.D. support. We also thank Benoit Chimier, Andreï Kabashin, Jürgen Reif, Matthieu Guillermin, and Razvan Stoian for fruitful discussions.

### References

[1]   A. Serpengüzel, A. Kurt, I. Inanç, J. E. Carey, E. Mazur, Journal Of Nanophotonics, **2**, 021770 (2008).

[2]   T. Sarnet, M. Halbwax, R. Torres, P. Delaporte, M. Sentis, S. Martinuzzi, V. Vervisch, F. Torregrosa, H. Etienne, L. Roux, S. Bastide, Proceedings SPIE, **6881**, 688119 (2008).

[3]   C. Wu, C. H. Crouch, L. Zhao, J. E. Carey, R. Younkin, J. A. Levinson, E. Mazur, R. M. Farrell, P. Gothoskar, A. Karger, Applied Physics Letters **78**(13) (2001).

[4]   A. Y. Vorobyev, C. Guo, Journal of Applied Physics **104**, 053516 (2008).

[5]   O. Varlamova, F. Costache, J. Reif, M. Bestehorn, Applied Surface Science **252**, 4702 (2006).

[6]   R. Belli, L. Toniutti, A. Miotello, P. Mosaner, D. Avi, Applied Physics A **92**, 217 (2008).

[7]   J. Reif, F. Costache, M. Henyk, S.V. Pandelov, Applied Surface Science **197-198**, 891 (2002).

[8]   M. Birnbaum, Journal of Applied Physics **36**, 3688 (1965).

[9]   J. Bonse, S. Baudach, J. Krüger, W. Kautek, M. Lenzner, Applied Physics A **74**, 19 (2002).

[10]   M. Csete, Zs. Bor, Applied Surface Science **133**, 5 (1998).

[11]   H. M. Van Driel, J. E. Sipe, J. F. Young, Physical Review Letters, **49**(26), 1955 (1982).

[12]   V. Zorba, N. Boukos, I. Zergioti, C. Fotakis, Applied Optics **47**, No. 11 (10 April 2008).

[13]   J. E. Sipe, J. F. Young, J. S. Preston, H. M. Van Driel, Physical Review B **27**(2), 1141 (1983).

[14]   N. S. Murthy, R. D. Prabhu, J. J. Martin, L. Zhou, R. L. Headrick, Journal of Applied Physics **100**, 023538 (2006).

[15]   J. Bonse, M. Munz, H. Sturm, Journal of Applied Physics **97**, 013538 (2005).

[16]   R. Taylor, C. Hnatovsky, E. Simova, Laser and Photonic Review **2**(1-2), 26 (2008).

[17]   S. I. Anisimov, V. A. Khokhlov, Instabilities in Laser-Matter Interaction, CRC Press, Boca Raton, FL (1995).

[18]   P. P. Rajeev, M. Gertsvolf, C. Hnatovsky, E. Simova, R. S. Taylor, P. B. Corkum, D. M. Rayner, V. R. Bhardwaj, Journal of Physics B: Atomic, Molecular and Optical Physics **40**, S273 (2007).

[19]   S. I. Anisimov, B.S. Luk'yanchuk, Physics Uspekhi, **45**(3), 293 (2002).

[20]   N. M. Bulgakova, R. Stoian, A. Rosenfeld, I. V. Hertel, W. Marine, E.E.B. Campbell, Applied Physics A **81**, 345 (2005).

[21]   H. M. Van Driel, Physical Review B **35**(15), 8166 (1987).

[22]   R. E. Simpson, Introductory Electronics for Scientists and Engineers, 2$^{nd}$ Edition. Allyn and Bacon (1987).

[23]   T. E. Itina, M. Mamatkulov, M. Sentis, Optical Engineering, **44**, 051109 (2005).

___________________________

[*]Corresponding author: thibault.derrien@lp3.univ-mrs.fr